\documentclass[english, a4paper,12pt]{article}

\usepackage[bookmarks, colorlinks=true, allcolors=blue, pagebackref=true, hyperfootnotes=false, pdfusetitle]{hyperref}
\usepackage[eng]{felipito}
\usepackage[round]{natbib}

\graphicspath{{./Graphics/}}

\spacing{1.2}

\author{Felipe Del Canto M.\footnote{Instituto de Economía, Pontificia Universidad Católica de Chile. Email : \href{mailto:fndelcanto@uc.cl}{fndelcanto@uc.cl}}}
\title{A complex net of intertwined complements: \\ Measuring interdimensional dependence among the poor}
\date{June 13, 2019}

\begin{document}

\maketitle
\thispagestyle{empty}

\vfill
\abstract{The choice of appropriate measures of deprivation, identification and aggregation of poverty has been a challenge for many years. The works of Sen, Atkinson and others have been the cornerstone for most of the literature on poverty measuring. Recent contributions have focused in what we now know as multidimensional poverty measuring. Current aggregation and identification measures for multidimensional poverty make the implicit assumption that dimensions are independent of each other, thus ignoring the natural dependence between them. In this article a variant of the usual method of deprivation measuring is presented. It allows the existence of the forementioned connections by drawing from geometric and networking notions. This new methodology relies on previous identification and aggregation methods, but with small modifications to prevent arbitrary manipulations. It is also proved that this measure still complies with the axiomatic framework of its predecessor. Moreover, the general form of latter can be considered a particular case of this new measure, although this identification is not unique.}
\vfill

\spacing{1.5}

\newpage
\section{Introduction}
The choice of appropriate measures of deprivation, identification and aggregation of poverty has challenged policymakers and researchers alike. The axiomatic framework developed by \cite{Sen76} was the tip of the iceberg for a rich literature in poverty measurement and the different properties they satisfy.\footnote{See, for example, \cite{Atkinson87}, \cite{FGTpaper}, \cite{Sen76, Sen79}, \cite{RavallionMPIBook}.} In the words of Lambert: ``Sen's article led to an industry of subsequent research, and has been cited by almost all subsequent writers as the cornerstone for measurement theory in this area''.\footnote{\cite{LambertPoverty}, p. 133.} Although focused primarily on unidimensional poverty measures, in \cite{LambertPoverty} the author summarizes the methodological issues associated with poverty measurement, presents the most common measures and discuss their relation with inequality and welfare. 

Right now, \textit{multidimensional poverty} is the one that has captured most of the attention thanks in part to the work of Alkire and Foster who introduced a robust framework to measure it and for which several desirable properties hold.\footnote{See \cite{AlkireFoster11} footnote 3 for a brief description of previous efforts in the literature to measure multidimensional poverty.} Although useful, one of the caveats of their approach is that it implicitly assumes different dimensions to be independent of each other. While simple in its application, it surely ignores the intrinsic dependence between different aspects of human life, just as Sen mentions: ``The (functionings) approach is based on a view of living as a combination of various ‘doings and beings’, with quality of life to be assessed in terms of the capability to achieve valuable functionings''.\footnote{\cite{SenFunctionings}.} 

In this work I present a variant of the usual method to measure deprivations that uses geometric and networking ideas to define interactions between dimensions. The basic notion, drawn from the work by \cite{BizarrePaper}, says that some dimensions are connected by ``paths'' that resemble the complementarity between them. This allows to redefine the deprivation gaps and with them both the identification and aggregation methods. For the former, I maintain the dual cutoff method of \cite{AlkireFoster11}, discussing briefly the minor adjustments in the choice of the second cutoff that are needed in this new setting. For aggregation, the adjusted $FGT$ class introduced in the same article must be adapted to correctly control for the size of the dependence between dimensions, so the aggregate poverty measure is not prone to manipulation by arbitrarily creating or vanishing connections. This \textit{network-adjusted} class of functions still satisfies the usual axioms, a fact that is proved in the main result of this article. The proof of the theorem also delivers one last insight regarding the choice of the interdimensional dependence structure: some of these choices can be interpreted as an implicit weight structure. This suggest an alternative way of defining weights that does not rely on determining the relative importance of each dimension directly, although the price to pay is that the number of parameters to be set or estimated grows quadratically with the number of dimensions. Finally, as is usual in the multidimensional measure of poverty, this alternative definition allows the use of weights although the correct definitions must be, again, done with care if we want the $FGT$ functions to still conform with the axiomatic framework previously mentioned.

The rest of the article is structured as follows. In \cref{sec:motivation} I give the main insights that support the definitions I present in \cref{sec:measuring}. In \cref{sec:identification} I discuss the use of the dual cutoff method for identification in this setting and redefine the $FGT$ class of functions to comply with it. In \cref{sec:weights} I outline how weights should be applied when the dependence structure is present. Finally, in \cref{sec:properties} I prove the main theorem of the article and discuss briefly its implications. In the final section I present some final remarks.

\section{Motivation} \label{sec:motivation}
\begin{figure}
		\centering
		\caption{The picture shows two districts from the 109th Congress in the US. Taken from Chambers and Miller (2010). The points and letters were added.}
		\subfloat[Sixth District, Maryland]{\label{fig:bizarre1} \includegraphics[width=0.3\textwidth]{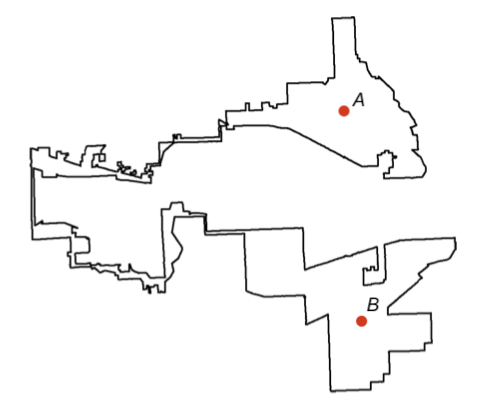}} \hspace{2ex}
		\subfloat[Fourth District, Illinois]{\label{fig:bizarre2} \includegraphics[width=0.3\textwidth]{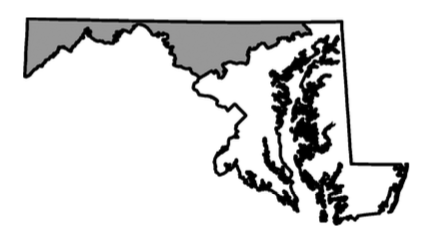}}
		\label{fig:bizarre}
	\end{figure}

\cite{BizarrePaper} propose a particular measure to determine what they call the ``bizarreness'' of the electoral districts in the US. The main insight is that a better district is capable of connecting more pairs of points by shorter paths without leaving the boundaries of it. For example, consider the situation depicted in \cref{fig:bizarre1}. In order to get from point $A$ to point $B$ (without leaving the borders) it is necessary to circumvent the whole district, when the ``logical'' way would be to just follow the straight line between those points. This makes a district worse than the one in \cref{fig:bizarre2}, whose paths between pairs points are much shorter than in the previous case.

\begin{figure}
		\centering
		\caption{Regular pentagon depicting the measure of deprivations in five dimensions of poverty, where $d_{i}$ represents dimension $i$ and the distance from $d_{i}$ to the centroid is the deprivation of that dimension.}
		\subfloat[Base case]{\label{fig:roads1} \includegraphics[width=0.3\textwidth]{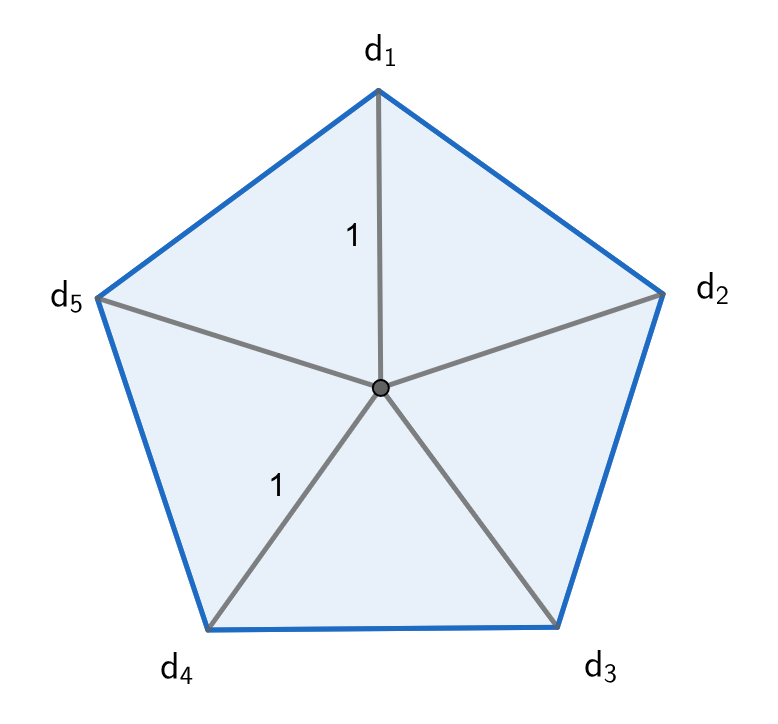}} \quad
		\subfloat[Deepened deprivation in $d_{1}$]{\label{fig:roads2}\includegraphics[width=0.3\textwidth]{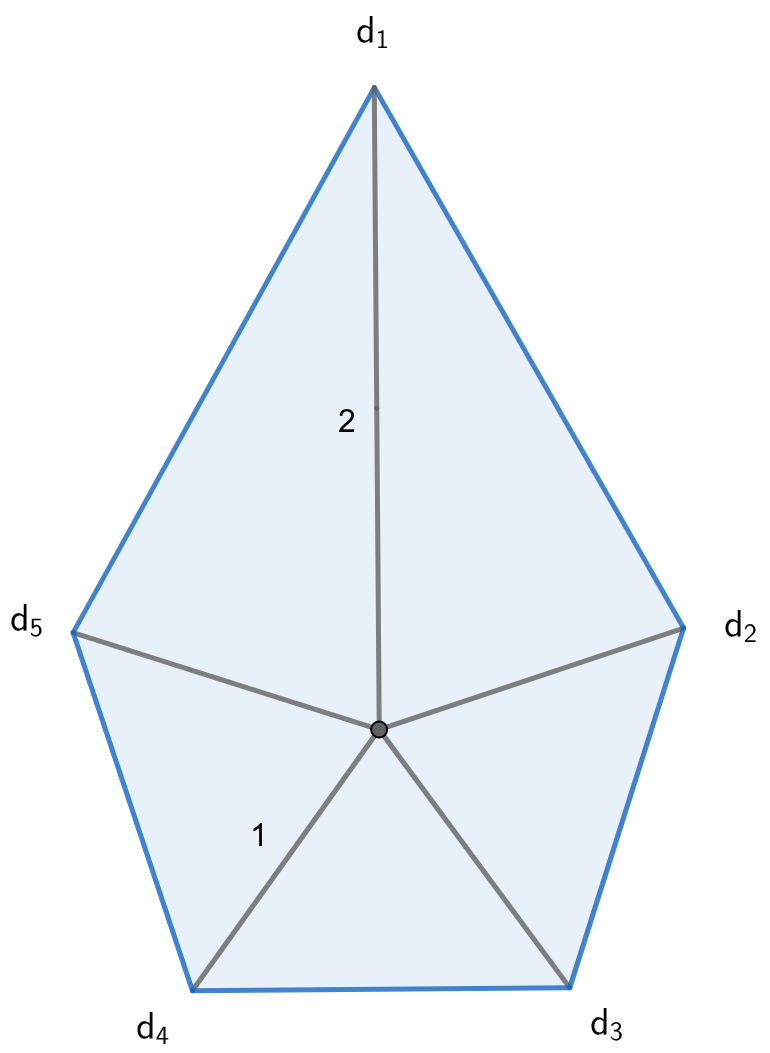}}
		
		\label{fig:roads}
	\end{figure}

Thus, drawing on this rationale, I take a geometric approach to the problem of measuring and interpreting the deprivation in each dimension of poverty that goes as follows. Suppose we are measuring five dimensions of poverty and that each of the five deprivation gaps for one individual is equal to 1. This scenario can be depicted using a regular pentagon, where each vertex corresponds to a different dimension and the distances from each vertex to the centroid are equal to 1, just as depicted in \cref{fig:roads1}. The basic notion is that in order to get from $d_{1}$ to $d_{2}$ -- which, for example, can be interpreted as employing health capabilities to obtain better achievements in education -- the individual must ``walk'' from the first vertex to the second passing through the centroid and in this case this implies a total walk length of 2. Imagine now that this person deepens its deprivation in $d_{1}$ (the health dimension) from 1 to 2. This of course negatively impacts its ability to employ health to obtain education, that is, she gets more deprived not only in $d_{1}$, but also in $d_{2}$ since the new length between both dimensions is equal to 3. This argument works the same way for the rest of the dimensions.

Of course, not every dimension should affect the other the same way. Following the previous analogy, health is an important condition to get better (achievements in) education but the converse does not need to be true at all or at least the education requirements needed to achieve certain goals in the health dimension can be radically lower. Using the language of \cref{fig:roads}, the previous observation implies that the length of the path when measured from $d_{1}$ to $d_{2}$ can differ from the length of the same path but when measured in the opposite direction. 

\section{Measuring deprivations} \label{sec:measuring}
Let $N$ be the number of people and $d$ the number of (poverty) dimensions. Let $i$ denote the index for individuals and $j$ the index for dimensions. For $\alpha \geq 0$ define,
	$$r_{ij}^{\alpha}:= \left(\frac{z_{j} - y_{ij}}{z_{j}}\right)^{\alpha} \uno[y_{ij} \leq z_{j}] \quad i \in \{1,\ldots,N\},\, j \in \{1,\ldots,d\},$$
where $y_{ij}$ denotes the achievement of person $i$ in dimension $j$, $z_{j}$ represents the $j$-dimension cutoff and $\uno[x]$ is the indicator function, which is equal to 1 if the argument is true and 0 otherwise. Recalling the interpretation in the previous section, note that $r_{ij}^{1}$ is analogous to the distance between the vertex $d_{j}$ and the centroid of the $d$-gon in \cref{fig:roads}.\footnote{However, since this definition forces $r_{ij}^{1} \in [0,1]$ the comparison between \cref{fig:roads1} and \cref{fig:roads2} no longer applies. Nevertheless, greater values of $r_{ij}^{1}$ do imply greater deprivations, which is the most important feature of the argument in the previous section.} 
Considering a wider range of values of $\alpha$ (instead of just $\alpha = 0$ or $1$) comes from the work of \cite{FGTpaper} and their $FGT$ measures. The principle behind this is that greater values of $\alpha$ (above 1) are more sensitive to minor changes in more deepeened deprivations, because the increase in $r_{ij}^{\alpha}$ given by a decrease in $y_{ij}$ is increasing as $y_{ij}$ goes to 0. On the other hand, values of $\alpha$ between 0 and 1 are much less sensitive to small changes, but penalize heavily. In the limit when $\alpha$ goes to 0, $r_{ij}^{\alpha}$ tends to 1, which means that even the smallest deprivations are given the maximum possible value, not allowing greater differentiation of the severity of poverty on each dimension.

Continuing with the argument of the previous section, $r_{ij}$ is not enough to measure the extent in which $i$ is deprived in the dimension $j$. We must incorporate the effect that the other $d-1$ dimensions have over $j$ and, as mentioned in the last part of \cref{sec:motivation}, we must allow these effects to vary between dimensions\footnote{A natural extension, which I do not explore here, is allowing the effects to vary also between individuals. Although this seems reasonable from a behavioral point of view, there are at least two concerns with this approach: 1) when $N$ increases (a common issue given poverty measuring is usually done at the country level), the number of parameters to set or estimate its too high to be reliable or even feasible and, 2) making the deprivation measure different for each individual is prone to ignore the aspects that affect a group or the population and as a whole.}. Let $M$ be a $d\times d$ matrix whose $jj'$ entry is $M_{jj'}$ and define
	\begin{equation}
		D_{ij}^{\alpha}
			:=	\frac{1}{d-1} \sum_{\substack{j'=1 \\ j' \neq j}}^{d} r_{ij}^{\alpha} + M_{jj'}r_{ij'}^{\alpha} \label{eq:Dalpha}.
	\end{equation}

Again, note that if $M_{jj'} = 1$ for every $j'$, then for $\alpha = 1$, $D_{ij}^{1}$ is just the average length of the $d-1$ paths from $d_{j}$ to $d_{j'}$ that pass through the centroid of the $d$-gon in \cref{fig:roads}. However, when not all entries of $M$ are equal, the effect of other dimensions over $j$ is heterogeneous. For future reference, observe that the $j$th row (resp. column) of $M$ tells how much dimension $j$ is affected by (resp. affects) the others. In what follows I will make the following assumptions. 

	\vspace{1ex}
	\qquad
	\textit{Assumption 1.} $M \in [0,1]^{d \times d}$, that is, $0 \leq M_{jj'} \leq 1$ for every $j,j' \in \{1, \ldots, d\}$.
	
	\qquad
	\textit{Assumption 2.} $M_{jj} = 1$ for all $j \in \{1, \ldots, d\}$
	\vspace{1ex}

Assumption 1 reflects two beliefs on the interdependence between dimensions. First, all the effects are nonnegative, that is, a higher deprivation in one dimension cannot reduce poverty in another. Moreover, it further implies that a pair of dimensions are not allowed to be substitutes.\footnote{As a final remark, note that this part of the assumption can be regarded as a variant of the monotonicity axiom, where increments in achievements in dimension $j$ (may) reduce poverty in another.} Second, the effect a dimension has over another cannot excede the effect it has on itself. The notion is that dependence between two dimensions is merely auxiliar, that is, it reflects the need of the individual to employ capabilities of one to obtain achievements in the other and not an intrinsic valuation of the former, something that should be captured by its own $D_{ij}$.\footnote{As a byproduct of this assumption note that $D_{ij}^{\alpha}$ is restricted between 0 and 2 (albeit not every one can reach 2), somehow reflecting the idea previously captured by $r_{j}^{\alpha}$ where the complete deprivation is coded with a 1 and the complete absence of it with a 0. In this case, complete deprivation in dimension $j$ implies that $D_{j}^{\alpha} \geq 1$, because poverty in other dimensions are allowed to increase this number above 1. In the other extreme, achievements above cutoff in dimension $j$ (i.e. $r_{j}^{\alpha} = 0$) can be coded with $D_{j}^{\alpha} > 0$, enforcing idea of complementarity between functionings.} Assumption 2 is harmless and can be regarded as an artifact to interpret $M_{jj'}$ as the relative effect of dimension $j'$ on $j$. Indeed, suppose that instead we had defined $D_{j}^{\alpha}$ as
	$$D_{ij}^{\alpha} = \frac{1}{d-1} \sum_{\substack{j'=1 \\ j' \neq j}}^{d} M_{jj} r_{ij}^{\alpha} + M_{jj'}r_{ij'}^{\alpha}.$$

Since $M_{jj}$ is constant for every term in the summation, then we can divide by $M_{jj}$ to obtain
	$$\frac{D_{ij}^{\alpha}}{M_{jj}} = \sum_{\substack{j'=1 \\ j' \neq j}}^{d} r_{ij}^{\alpha} + \frac{M_{jj'}}{M_{jj}}r_{ij'}^{\alpha}. $$

If $M_{jj}$ is interpreted as the intrinsic valuation that society gives to dimension $j$, then the ratio $\frac{M_{jj'}}{M_{jj}}$ is precisely the relative effect that $j'$ has on $j$ measured in the terms of the latter. Of course, this motivates considering the $M_{jj'}$ as prices the society defines to pay to ``interchange poverty'' between dimensions $j$ and $j'$ and in that interpretation Assumption 2 is nothing more than a regularization drawing from general equilibrium theory saying that what matters is relative and not absolute prices.\footnote{See for example \cite{JehleReny}, p. 211} 

In order to make the rest of the discussion clearer, I will make the following definitions. The matrix $M$ and the effects between dimensions will be called the dependence structure. For $j, j' \in \{1, \ldots, d\}$, we say that $j'$ is connected to $j$ if $M_{jj'} > 0$. If both $M_{jj'}, M_{j'j} > 0$, then we just say that $j$ and $j'$ are connected. For a given dimension $j$, the set $C_{j} := \{j' \in \{1, \ldots, d\} : M_{jj'} > 0\}$ and the dimensions inside it are called the connections of $j$. If for every $j$ we have $C_{j} = \{j\}$ then the dependence structure is said to be disconnected.

Back to equation (\ref{eq:Dalpha}), a quick inspection of it should convince the reader that
	\begin{equation}
		D_{ij}^{\alpha}
			=	r_{ij}^{\alpha} + \frac{1}{d-1} \sum_{\substack{j'=1 \\ j' \neq j}}^{d} M_{jj'}r_{ij'}^{\alpha} \label{eq:Dalpha2}.
	\end{equation}

Of course, if $M = \mathbf{I}_{d}$, where $\mathbf{I}_{d}$ is the identity $d \times d$ matrix, then $D_{j}^{\alpha} = r_{j}^{\alpha}$ and we retrieve the usual metric for deprivations. \cref{eq:Dalpha2} is easier to compute and is cleaner than its predecessor and will be the one I will use in the rest of the document. As a final observation about $D_{j}^{\alpha}$, suppose $\alpha > 0$ and note that if we define $\eta_{ij} := y_{ij}/z_{j}$, then
	\begin{equation} \label{eq:dDdy}
		\dfdx{D_{ij}^{\alpha}}{\eta_{ij'}}
			=	\left\{ \begin{aligned}
					&\dfdx{r_{ij}^{\alpha}}{\eta_{ij}}				&&\text{if } j' = j,	\\
					&\frac{M_{jj'}}{d-1}\dfdx{r_{ij'}^{\alpha}}{\eta_{ij'}}	&&\text{otherwise}
				\end{aligned} \right. .\footnote{Where all these derivatives must be understood for $0 < \eta_{ij} < 1$.}
	\end{equation}
But since $r_{ij}^{\alpha}$ is defined in the same way for every $j$, then
	$$\theta := \dfdx{r_{ij}^{\alpha}}{\eta_{ij}} = \dfdx{r_{ij'}^{\alpha}}{\eta_{ij'}}, \qquad \paratodo j, j' \in \{1, \ldots, d\},$$
and we can rewrite equation (\ref{eq:dDdy}) as
	$$\dfdx{D_{ij}^{\alpha}}{\eta_{j'}}
			=	\left\{ \begin{aligned}
					&\theta					&&\text{if } j' = j,	\\
					\frac{M_{jj'}}{d-1}\,&\theta	&&\text{otherwise}
				\end{aligned} \right. .$$

Again, from the definition of $r_{ij}^{\alpha}$ we know that $\theta < 0$ and from Assumption 1, 
	$$0 \leq \frac{M_{jj'}}{d-1} \leq \frac{1}{d-1}.$$

Thus, $\dfdx{D_{ij}^{\alpha}}{\eta_{ij'}}$ is nonpositive if $j' \neq j$ and strictly negative if $j' = j$, preserving the desirable properties of $r_{j}^{\alpha}$, that is, that better achievements lead to less poverty. Finally, if all dimensions apart from $j$ increase their relative achievements by the same amount, then the total effect on $D_{ij}^{\alpha}$ is
	\begin{equation} \label{eq:totaleffect}
		\frac{\theta}{d-1} \sum_{\substack{j' = 1 \\ j' \neq j}}^{d} M_{jj'},
	\end{equation}
which by Assumption 1 lies between 0 and $\theta$. In the extreme case where the summation in equation (\ref{eq:totaleffect}) is equal to $d-1$ (when $M = \mathbf{I}_{d}$), then the relative increase on all other dimensions equals the effect of the increase in that dimension, reflecting that complementarities do not work in a one-to-one fashion. Instead, for the interdimensional dependence to match the own effects a joint improvement is necessary.

\section{Identification and measuring of poverty} \label{sec:identification}
In order to proceed with identification a few more definitions are in place. Let $D^{\alpha}$ be the $N \times d$ matrix whose $ij$ entry is $D_{ij}^{\alpha}$. Also define
	$$D_{i} := \sum_{j=1}^{d} D_{ij}^{0}.$$
Again, if $M = \mathbf{I}_{d}$, then $D_{i}$ es equal to the deprivation count $c_{i}$ presented in \cite{AlkireFoster11}. However, in the general case, $D_{i}$ will be greater than just the number of deprivations. Calling $\Sigma$ the sum of all elements of matrix $M$ and $\Sigma_{j}$ the sum of its $j$th column, the reader will be convinced that using the definition of $D_{ij}^{0}$ and a little manipulation the following lemma holds.

\begin{lem} \label{lem:bounds} The highest value of $D_{i}$ is
	\begin{equation} \label{eq:upperbound}
		\overline{d} := d + \frac{\Sigma - d}{d-1}.
	\end{equation}
Whereas the lowest (non-zero) possible value is
	\begin{equation} \label{eq:lowerbound}
		\underline{d} := 1 + \frac{\min_{1 \leq j \leq d} \Sigma_{j} - 1}{d-1},
	\end{equation}
\end{lem}

\begin{proof} The upper bound is reached when $r_{ij}^{0} = 1$ for every $j$. Thus, using equation (\ref{eq:Dalpha2}) we obtain 
	\begin{align*}
		\overline{d}
			&=	d + \frac{1}{d-1}\sum_{j=1}^{d} \sum_{\substack{j' = 1 \\ j' \neq j}}^{d} M_{jj'}
			=	d + \frac{1}{d-1}\sum_{j=1}^{d} \sum_{j' = 1}^{d} M_{jj'} - \underbrace{M_{jj}}_{= 1}
			=	d + \frac{1}{d-1} \left(\Sigma - d \right),
	\end{align*}

which is equation (\ref{eq:upperbound}). The lower (non-zero) bound is obtained when only one dimension has $r_{ij}^{0} = 1$. Since $r_{ij}^{0}$ appears in every $D_{ij'}^{0}$, then using that $r_{ij'}^{0} = 0$ if $j' \neq j$ we have
	\begin{align*}
		D_{i}	&=	1 + \frac{1}{d-1} \sum_{\substack{j' =1 \\ j' \neq j}}^{d} M_{j'j}
			=	1 + \frac{1}{d-1} \left(\sum_{j'=1}^{d} M_{j'j} - M_{jj} \right)
			=	1 + \frac{\Sigma_{j} - 1}{d-1},
	\end{align*}
which is minimized when $\Sigma_{j}$ takes its lower value, giving equation (\ref{eq:lowerbound}).
\end{proof}

The previous lemma states that the highest and lowest possible values of $D_{i}$ depend not only $d$ but also on the structure of the interdimensional dependence. It should come with no surprise that when $M = \mathbf{1}_{d \times d}$ (the $d \times d$ matrix full of ones) then $\overline{d} = 2d$, which is exactly $d$ times the highest value of $D_{ij}^{0}$. Likewise, when $M = \mathbf{I}_{d}$, then $\underline{d} = 1$ and $\overline{d} = d$ which is the standard lower and upper bounds for the deprivation count $c_{i}$. In general, the bounds on $D_{i}$ depend on a normalized ``distance'' between $M$ and $\mathbf{I}_{d}$: the second term in both equations \eqref{eq:upperbound} and \eqref{eq:lowerbound}. To see this, let's fix our attention on $\overline{d}$ and note that when $M = \mathbf{I}_{d}$, then $\Sigma = d$ and the distance is 0. As the remaining entries of $M$ grow from 0 to 1 (i.e. moving away from $\mathbf{I}_{d}$), the difference $\Sigma - d$ increases until $M = \mathbf{1}_{d\times d}$, where $\Sigma - d = d^{2} - d$ and the distance equals $d$. As for $\underline{d}$, this normalized distance is more local and compares the ``minimum column'' of $M$ and $\mathbf{I}_{d}$.\footnote{Where minimum column must be understood as the one that satisfies $\Sigma_{j} = \min_{1 \leq j' \leq d} \Sigma_{j'}$. Note that for $\mathbf{I}_{d}$, all columns sum the same.} In terms of this setting, this distance relates how much the minimum total dependence exerted by one dimension over the others compares with not affecting at all. 

The last ingredient needed to proceed with identification is how $D_{i}$ changes when for some $j$, $r_{ij}^{0}$ changes from 0 to 1. Recall that the analogue of $D_{i}$, $c_{i}$, increases by 1 every time the achievements in one dimension get below their cutoff. The jump in $D_{i}$ is potentially bigger if dependence structures are present. Specifically, if $M \neq \mathbf{I}_{d}$ then, following the calculations of \cref{lem:bounds}, it is exactly equal to $\delta_{j} := 1 + \frac{\Sigma_{j} - 1}{d-1}$. Without loss of generality, suppose that the columns of $M$ are arranged in ascending order of $\Sigma_{j}$ and thus $\delta_{1} \leq \delta_{2} \leq \cdots \leq \delta_{d}$. Suppose we start with $D_{i} = 0$ (that is, $r_{ij}^{0} = 0$ for every $j$) and observe that $D_{i} \geq \delta_{j}$ if at least one $r_{ij'}^{0}$ changes from 0 to 1 for $j' \geq j$. Evidently, the converse is not necessarily true. Note that if $\delta_{1} + \delta_{2} \geq \delta_{3}$, then changing $r_{i1}^{0}$ and $r_{i2}^{0}$ from 0 to 1 also implies $D_{i} \geq \delta_{3}$. Thus, the ordering of the $\delta_{j}$ is not the only feature that matters but also where the sum of a subset of the $\delta_{j}$ is placed in that disposition.

Now we are ready to describe de identification procedure. For this purpose I will rely on the \textit{dual cutoff} method of \cite{AlkireFoster11}. Let $\underline{d} \leq k \leq \overline{d}$ be any real number and let $y_{i}$ and $z$ be the $d$-dimensional vectors of achievements and cutoffs, respectively. Define
	$$
		\rho_{k}(y_{i}; z)
			:=	\left\{ \begin{aligned}
						&1	&\text{if } D_{i} \geq k	\\
						&0	&\text{otherwise}
					\end{aligned} \right. .
	$$
This function has the usual task of identifying if person $i$ is poor or not. The previous discussion about the discontinuities in the value of $D_{i}$ showed that, although the jumps are discrete and can be thoroughly described, the use of a continuous $k$ facilitates the discussion.\footnote{In this sense, $\rho_{k}$ so defined works in a similar fashion as the one defined in \cite{AlkireFoster11} when using weights. Consequently in \cref{sec:weights}, when I describe the use of weights in this setting, this issue about the nature of $k$ will be meaningless.} Observe that since we are following the same identifying function as in Alkire and Foster, then its desirable properties still hold: 1) it is `poverty focused' because increases in any achievement reduces (or leaves unchanged) the value of $D_{i}$, thus leaving $\rho_{k}$ unchanged if it was already at 0 and, 2) it is `deprivation focused' because if $y_{ij} \geq z_{j}$ for some $i,j$, then $r_{ij}^{0} = 0$ and thereby increases in $y_{ij}$ leave $r_{ij}^{0}$, $D_{i}$ and, of course, $\rho_{k}$ unchanged. The reader should note that, even when considering interdimensional dependence, the identification method remained the same as in the disconnected case. The reason is that this approach only changes the way deprivations are measured (that is, the values of $D_{ij}^{0}$), stating that it depends on the achievements on its connections. Since the dual cutoff method is already measuring poverty based on deprivations and has the two desirable properties mentioned before, all necessity of a different strategy disappears. 

Continuing with aggregation, let $y$ be the $N \times d$ matrix of achievements. In presence of dependence structures, the usual method based in the adjusted $FGT$ class needs a minor change. To see this, observe that using the standard definition we would have
	\begin{equation} \label{eq:wrongFGT}
		FGT_{\alpha}(y ; z) := \frac{1}{Nd} \sum_{(i,j)} D^{\alpha}_{ij} \cdot \rho_{k}(y_{i} ; z), \qquad \alpha \geq 0.
	\end{equation}
Since the $D_{ij}^{\alpha}$ now incorporate a network component, this simple average is uninformative, because $FGT_{\alpha}$ can be artificially modified by harnessing the monotonicity with respect to the dependence structure: including connections will necessarily increase it and removing them will have the opposite effect. In order to overcome this issue, it is necessary that the denominator controls for the ``size'' of the dependence structure. A natural way to implement this correction is to replace the denominator in equation (\ref{eq:wrongFGT}) by the highest possible value of the summation, $N\overline{d}$.\footnote{To see this, note that for fixed $i$, the sum over $j$ attains its maximum value when $\rho_{k}(y_{i}; z)$ and every $D_{ij}^{\alpha}$ are equal to 1, and by \cref{lem:bounds} this value is equal to $\overline{d}$. The result thus follows from summing $\overline{d}$ over $i$.} In consequence, consider the family of \textit{network}-adjusted $FGT$ functions defined by
	\begin{equation} \label{eq:rightFGT}
		FGT_{\alpha}(y ; z) := \frac{1}{N\overline{d}} \sum_{(i,j)} D^{\alpha}_{ij} \cdot \rho_{k}(y_{i} ; z), \qquad \alpha \geq 0.
	\end{equation}
This family retains the desirable properties of its precursor, something I will prove in \cref{sec:properties}. Opposed to previous literature, in presence of dependence structures the (multidimensional poverty measuring) methodology now is a triple composed of an identification method, an aggregate measure and the dependence structure. That is, the methodology with structure $M$, dual cutoff identification and $FGT$ aggregation can be written as the triple $\clx{M}_{k,\alpha, M} := (\rho_{k}, FGT_{\alpha}, M)$.

\section{Weights} \label{sec:weights}
In the previous definitions, while there is a dependence structure, all dimensions are considered of equal importance when computing $D_{i}$. This is of course not desirable in applications, where different societies may have different choices of the relative importance of each dimension. This point is remarked in \cite{AlkireFoster11} in the words of both Atkinson and Sen. Let $w$ be a $d$-dimensional vector of positive real values where each coordinate represents the weight associated with the corresponding dimension.\footnote{As opposed to the common ``weight'' assumption that the sum of all of them is 1. In this case we will assume this sum is equal to the number of dimensions, $d$.} The usual way to implement weights is to redefine $r_{ij}^{\alpha}$ as $w_{j}r_{ij}^{\alpha}$ and continue with identification and aggregation in a similar fashion as the one described in the previous section. Observe that in this setting, following this path will lead to apply weights for every dimension when computing the $D_{ij}^{\alpha}$, because we would have
	\begin{equation} \label{eq:badWeights}
		D_{ij}^{\alpha} = w_{j}r_{ij}^{\alpha} + \frac{1}{d-1} \sum_{\substack{j' = 1 \\ j' \neq j}}^{d} M_{jj'} w_{j'}r_{ij'}^{\alpha},
	\end{equation}
from where two concerns raise. First, weights are intended to highlight the relative importance of each dimension for identification and aggregation purposes, not for measuring deprivations, which is what is happening here by altering the value of $D_{ij}^{\alpha}$ before aggregating. Second, by looking closely at the summation in equation (\ref{eq:badWeights}), we note that each $M_{jj'}$ is multiplied by the weight $w_{j'}$. This can be seen as the weights altering the dependence structure when in fact both should be unrelated to each other, again on the basis that deprivation measuring is not connected with identification. Another way to see this last issue is: if we intentionally modify the dependence structure using weights, then the former was ill defined from the beginning and it should have been changed, no matter the choice of weights.

By the previous observation, the proper way of using weights is including them after computing $D_{ij}^{\alpha}$, that is, redefining $D_{ij}^{\alpha}$ as $w_{j}D_{ij}^{\alpha}$. Observe that all the definitions and calculations made in the previous sections implicitly considered $w_{j} = 1$ for every $j$. Additionally, note that
	$$D_{i} = \sum_{j=1}^{d} w_{j}D_{ij}^{0},$$
and since $w_{j}$ can be any positive real number (lower than $d$), the discussion in \cref{sec:identification} about the possible values of $D_{i}$ no longer holds. This implies that for identification purposes, the value of $k$ in $\rho_{k}(y_{i}; z)$ now ranges from 0 to 
	\begin{equation} \label{eq:upperBoundweights}
		\tilde{d} := d + \frac{1}{d-1}\left(\sum_{j=1}^{d} w_{j}\Sigma_{j} - d \right).
	\end{equation}
Indeed, since the maximum value for $D_{ij}^{0}$ is 
	$$1 + \frac{\Sigma_{j} - 1}{d-1},$$
then the maximum value for $D_{i}$ when using weights is
	\begin{align*}
		\tilde{d}
			&=	\sum_{j=1}^{d} w_{j} \left(1 + \frac{\Sigma_{j} - 1}{d-1} \right)	\\
			&=	\sum_{j=1}^{d} w_{j} + \frac{1}{d-1} \left( \sum_{j=1}^{d} w_{j}\Sigma_{j} - \sum_{j=1}^{d} w_{j} \right)	\\
			&=	d + \frac{1}{d-1}\left(\sum_{j=1}^{d} w_{j}\Sigma_{j} - d \right)
	\end{align*}
which is exactly equation (\ref{eq:upperBoundweights}). A similar analysis as the one in the previous section about the ``distance'' between $M$ and $\mathbf{I}_{d}$ could be made here. However, I restrain from developing it as the arguments are basically the same and no additional insight could be gained from it.

In terms of aggregation, the $FGT$ class is defined in a similar way as before, that is,
	\begin{equation} \label{eq:rightFGTweights}
		FGT_{\alpha}(y; z) = \frac{1}{N\tilde{d}} \sum_{(i,j)} w_{j}D_{ij}^{\alpha} \cdot \rho_{k}(y_{i}; z).
	\end{equation}
Again, note that the denominator corrects for the size of the network, just as in the previous section. Finally, note that when using weights, the methodology is not a triple but the 4-tuple $\clx{M}_{\alpha, k, M, w} = (\rho_{k}, FGT_{\alpha}, M, w)$.
	
\section{Properties} \label{sec:properties}
In this section I will prove that this new measure still satisfies the same properties as the disconnected case presented in \cite{AlkireFoster11}. For a complete description on the axiomatic framework of that article the reader is referred to \cref{sec:axioms}. For the sake of readability, I will call $\clx{M}$ to the methodology $\clx{M}_{\alpha, k, M, w}$.

\begin{teo} For any given vector of weights $w$, cutoffs $z$ and dependence structure $M$, the methodology $\clx{M}$ satisfies decomposability, replication invariance, symmetry, poverty and deprivation focus, weak and dimensional monotonicity, nontriviality, normalization, and weak rearrangement for $\alpha \geq 0$. Also, if $\alpha > 0$ it satisfies monotonicity and if $\alpha \geq 1$, weak transfer.
\end{teo}

\begin{proof} First, observe that since $M$ and $w$ are fixed, then $\tilde{d}$ does not change with $\alpha$. First, suppose that $\alpha \geq 0$. For decomposability note that by definition
	\begin{align*}
		FGT_{\alpha}(x,y; z)
			&=	\frac{1}{(n(x) + n(y))\tilde{d}} \left(\sum_{\substack{(i.j) \\ i \in y}} w_{j}D_{ij}^{\alpha}\rho_{k}(y_{i}, z_{i}) + \sum_{\substack{(i.j) \\ i \in x}} w_{j}D_{ij}^{\alpha}\rho_{k}(x_{i}, z_{i}) \right)	\\
			&=	\frac{n(x)}{n(x) + n(y)} \cdot \frac{1}{n(x)\tilde{d}} \sum_{\substack{(i.j) \\ i \in y}} w_{j}D_{ij}^{\alpha}\rho_{k}(y_{i}, z_{i})	\\
			&\qquad\qquad\qquad
					+ \frac{n(y)}{n(x) + n(y)} \cdot \frac{1}{n(y)\tilde{d}} \sum_{\substack{(i.j) \\ i \in x}} w_{j}D_{ij}^{\alpha}\rho_{k}(x_{i}, z_{i}),
	\end{align*}
proving the desired result. Replication invariance can be seen as a particular case of the previous axiom, by noting that if $x = (y ,\ldots, y)$ ($m$ times) then $x$ can be disaggregated into $m$ groups of size $N$. Thus, using decomposability
	$$FGT_{\alpha}(x;z) = \sum_{k=1}^{m} \frac{N}{mN} FGT_{\alpha}(y; z) = FGT_{\alpha}(y;z) \sum_{k=1}^{m} \frac{1}{m} = FGT_{\alpha}(y;z).$$

Symmetry follows directly from the associativity of the sum operation and by the definition of $FGT_{\alpha}$. Observe now that if $x$ is a simple increment of $y$ among the non-poor, then $\rho_{k} = 0$ for the person $i'$ whose achievement was increased. This means that all terms $D_{i'j}^{\alpha}$ for person $i'$ were not part of the summation in $FGT_{\alpha}(y;z)$ and nor are the new $D_{i'j}^{\alpha}$ in $FGT(x;z)$ because $\rho_{k}$ remains unchanged for person $i'$. Thus poverty focus is satisfied. Now note that if $i'$ is non-deprived in $j'$, then all the terms $r_{i'j'}^{\alpha}$ were equal to 0 and thus were not engaged in the calculation of any $D_{i'j}^{\alpha}$. Hence, increasing achievement $y_{i'j'}$ has no effect on any term in the summation and $FGT_{\alpha}$ remains unchanged, proving that $\clx{M}$ satisfies deprivation focus.

For the monotonicity axioms, note that a simple increment in coordinates $i'j'$ of matrix $y$ either decreases or leaves unchanged the value of every $D_{ij}^{\alpha}$, thus decreasing or leaving unchanged the value of $FGT_{\alpha}$, satisfying weak monotonicity. Now, if the matrix $x$ is obtained by a dimensional increment among the poor, then at least $r_{i'j'}^{\alpha}$ is strictly decreased from a positive value to 0, while the other terms either decrease or stay the same and this implies that $FGT_{\alpha}(x ; z) < FGT_{\alpha}(y ; z)$, proving that $\clx{M}$ satisfies dimensional monotonicity.

Nontriviality and normalization follow from observing that by definition $FGT_{\alpha}(0; z) = 1$ and $FGT_{\alpha}(z; z) = 0$, where the $z$ in the first argument is the matrix whose $ij$th entry is $z_{ij} = z_{j}$. Now suppose $x$ is obtained by an association decreasing rearrangement among the poor. This means that (without loss of generality) $y_{i} \geq y_{i'}$ in the vector dominance sense. After rearranging, person $i$ still remains poor, since her achievements either decreased (dimensions $j$ that were interchanged) or stayed the same (dimensions $j$ that were not interchanged). Person $i'$ also remains poor because, although her achievements increased, the fact that $y_{i} \geq y_{i'}$ implies these are still lower or equal as the ones $i$ had before the rearrangement, when this person was poor. Since $\rho_{k}$ does not change and the terms involved in the sum of $FGT_{\alpha}$ are just rearranged, then the weak rearrangement axiom is satisfied.  

Now suppose $\alpha > 0$ and that $x$ is a simple increment from $y$ among the poor. Since $\alpha \neq 0$, then $r_{i'j'}^{\alpha}$ strictly decreases, which in turn strictly decreases $FGT_{\alpha}$, because $\rho_{k}(y_{i}; z) = 1$.

Finally, for $\alpha \geq 1$, suppose we average achievements among the poor. Observe that a tedious but possible manipulation of equation (\ref{eq:rightFGTweights}) using the definition of $D_{ij}^{\alpha}$ leads to
	\begin{equation} \label{eq:FGTproof}
		FGT_{\alpha}(y; z) = \frac{1}{N\tilde{d}} \sum_{(i,j)} A_{j}r_{ij}^{\alpha}\rho_{k}(y_{i}; z),
	\end{equation}
where $A_{j}$ is a positive constant.\footnote{The exact value of $A_{j}$ is presented in equation (\ref{eq:Aj}). Although the precise expression is meaningless here, it regains importance in the discussion following the proof.} This shows that $FGT_{\alpha}$ is convex in $r_{ij}^{\alpha}$.\footnote{When $\alpha = 1$, $FGT_{\alpha}$ is linear in $r_{ij}^{\alpha}$. In that case all the convexity inequalities are satisfied with equality but the argument still holds.} If nobody changes their poverty status, then this convexity implies that averaging achievements leads to a lower value of $FGT_{\alpha}$. In the case where the poverty status of some individuals do change, then thos $r_{ij}^{\alpha}$ are taken out of the summation and again convexity leads the desired result. In sum, $\clx{M}$ verifies weak transfer for $\alpha \geq 1$.
\end{proof}

To close this section, I briefly discuss how a particular choice of dependence structure can lead to an undesirable result. The exact value of $A_{j}$ in equation (\ref{eq:FGTproof}) is
	\begin{equation} \label{eq:Aj}
		A_{j} = w_{j} + \frac{1}{d-1} \sum_{j' \neq j} M_{j'j}w_{j'}. \footnote{This quantity is indeed positive under the assumptions made so far.}
	\end{equation}
Thereby, we could rewrite that equation as follows
	$$FGT_{\alpha}(y; z) = \frac{1}{Nd} \sum_{(i,j)} \underbrace{\frac{d A_{j}}{\tilde{d}}}_{\tilde{A}_{j}}\;  r_{ij}^{\alpha}.$$ 
This expression looks strikingly similar as the usual $FGT_{\alpha}$ with weights. For this to be true, the sum of the $\tilde{A}_{j}$ must be equal to $d$ or, equivalently, the sum of the $A_{j}$ must be equal to $\tilde{d}$. A little manipulation allows to see that
	$$\sum_{j} A_{j}
		=	d +  \frac{1}{d-1} \sum_{j} \left(\sum_{j'} M_{j'j}w_{j'} - w_{j}\right)
		=	d + \frac{1}{d-1}\left(\sum_{j'} w_{j'} \sum_{j} M_{j'j} - d \right),$$
which is in turn very similar to $\tilde{d}$. The difference is that the term $\sum_{j} M_{j'j}$ is the sum of the $j'$th row while in $\tilde{d}$ that term corresponds to the sum of the $j$ column ($\Sigma_{j}$). This implies that by choosing a symmetric dependence structure -- that is, a symmetric matrix $M$ -- then $M_{jj'} = M_{j'j}$ and we would have
	$$\sum_{j} A_{j} = d + \frac{1}{d-1}\left(\sum_{j} w_{j}\Sigma_{j} - d\right) = \tilde{d},$$
showing that in this case we have an implicit weight choice. This result suggests that a possible way to specify weights is to start with a certain symmetric dependence structure and, of course, do not apply other set of weights. In that case, the implicit weights would be
	\begin{equation} \label{eq:impliedWeights}
		w_{j}		= \frac{d}{\overline{d}}\left(1 + \frac{1}{d-1} \sum_{j' \neq j} M_{jj'} \right)
				= d \cdot \frac{\left(1 + \frac{\Sigma_{j} - 1}{d-1} \right)}{\overline{d}}. \footnote{Note that since $\overline{d}$ is the sum of all the numerators over $j$, then the fraction is less than 1 and each $w_{j}$ is less than d.}
	\end{equation}
Although this method alleviates the concern of choosing ``the most important'' dimensions directly and replaces it with just choosing individually the connections for each dimension, it increases the number of parameters to choose from $d-1$ to $\frac{d(d-1)}{2} - 1$ (the number of elements above the diagonal of $M$ minus 1). In particular, when using implied weights, the parameters grow quadratically with the number of dimensions. Another possible interpretation is that all weighted schemes are just particular cases of the presence of dependence structures.\footnote{To maintain the discussion short I do not present the proof of this result. However, it follows directly from setting the system of equations given by equation \eqref{eq:impliedWeights} and noting that the matrix associated with the vector of $\Sigma_{j}$ is nonsingular.} However, as equation \eqref{eq:impliedWeights} suggest, for any given set of weighs there are several dependence structures that imply them, since the only important quantities are the $\Sigma_{j}$ and these can be equal for different (symmetric) dependence structures.

\section{Concluding remarks}
In this article I propose an alternative way to measure deprivations that allows the presence of dependence between dimensions. This measure draws notions from geometry and networking by interpreting dimensional complementarities as ``paths'' and deprivation gaps as the lengths of those paths. By using lightly modified identification and aggregation functions, the methodology developed not only allows the use of weights for representing the relative importance of each dimension of poverty but also conforms to the axiomatic framework used in the literature.

The proof of the main theorem of the paper shows that in particular cases the dependence structure implies a set of weights, suggesting an alternative way to derive them without setting the relative importance of each other but instead specifying the connections each dimension has with the others. The caveat in this alternate method is that the number of parameters increases quadratically with the number of dimensions. A final interpretation of this result is that it is possible to consider any set of weights as coming from a certain dependence structure, although this structure may not be unique.

An important point that has not been discussed here is the choice of the dependence matrix $M$. Of course, it could be arbitrarily fixed by the authority but it could be estimated instead. For example, to set the effect of health over education, a possible proxy can be the impact of truancy over school achievements. Estimates like this may find better support in the literature than particular choices of weights and could help not only reduce the number of parameters to be set but also diminish the possibility of artificially altering the poverty count.

As a final comment, the previous discussion assumed that all dimensions are complements. This was backed up by the idea that there is no replacement between dimensions or, in other words, that the choice of dimensions aggregates them just enough to be considered as different functionings. It is still possible to pursue the idea of substitutability, for example, by relaxing $M_{jj'}$ to take values in $[-1,1]$. In this case the choice of identification and aggregation functions must be donde with care in order to preserve desirable properties or even be well defined for any given dependence structure ($\tilde{d}$ could be 0 for some $M$ and $w$). Also observe that under this approach, a person decreasing her achievements in one deprived dimension could potentially increase her performance in another and, depending on the dependence structure and the identification method, this could also rise her out of poverty. This could be undesirable and extremely prone to manipulation if the poorest (in terms of income) groups in the society turn out to be non-deprived in some dimensions that could be exploited to reduce multidimensional poverty.

\newpage
\appendix
\renewcommand{\thesubsection}{\Alph{subsection}}

\section*{Appendix}

\subsection{Axiomatic Framework} \label{sec:axioms}
As in \cref{sec:properties}, the methodology $\clx{M}_{\alpha, k, M, w}$ will be called just $\clx{M}$.

	\vspace{2ex}
		\textsc{Decomposability.} Suppose we disaggregate the matrix of achievements into two matrices $x$ and $y$, representing 2 (not necessarily exhaustive) subgroups of sizes $n(x)$ and $n(y)$, respectively. Then $\clx{M}$ satisfies decomposability if
	$$FGT_{\alpha}(x,y ; z) = \frac{n(x)}{n(x) + n(y)} FGT_{\alpha}(x; z) + \frac{n(y)}{n(x) + n(y)} FGT_{\alpha}(y; z).$$ 
	
	\vspace{2ex}
		\textsc{Replication invariance.} If the matrix of achievements $x$ is obtained by merging the same matrix $y$ a certain number of times, then $x$ is called a replication of $y$. In this context, $\clx{M}$ satisfies replication invariance if $FGT_{\alpha}(x ; z) = FGT_{\alpha}(y ; z)$.

	\vspace{2ex}
		\textsc{Symmetry.} Suppose we obtain the matrix of achievements $x$ by rearranging the rows of achievement matrix $y$. Then we say that $x$ is a permutation of $y$ and further, if $FGT_{\alpha}(x ; z) = FGT_{\alpha}(y; z)$ then $\clx{M}$ is said to satisfy symmetry.

\begin{defn*}[Increments] We say that the matrix $x$ is obtained from $y$ by \textit{a simple increment} if for some entry $i'j'$, $x_{i'j'} > y_{i'j'}$ and for every other entry $ij$, $x_{ij} = y_{ij}$. Further, if $\rho_{k}(y_{i'}, z) = 0$ (that is $i'$ is not poor) then we say that it is a simple increment \textit{among the non-poor}. If $y_{i'j'} > z_{j'}$, then it is a simple increment \textit{among the non-deprived}. We call the simple increment, \textit{deprived increment among the poor} if $z_{j'} > y_{i'j'}$ and $i'$ is poor. Finally, the simple increment is called \textit{dimensional increment among the poor} if in addition to be a deprived increment among the poor, we have $x_{i'j'} > z_{j'}$.
\end{defn*}

	\vspace{2ex}
		\textsc{Poverty Focus.} If $x$ is obtained from $y$ by a simple increment among the non-poor, then $FGT_{\alpha}(x ; z) = FGT_{\alpha}(y; z)$.

	\vspace{2ex}
		\textsc{Deprivation Focus.} If $x$ is obtained from $y$ by a simple increment among the non-deprived, then $FGT_{\alpha}(x ; z) = FGT_{\alpha}(y; z)$.

	\vspace{2ex}
		\textsc{Weak monotonicity.} $\clx{M}$ satisfies weak monotonicity if whenever $x$ is obtained from $y$ by a simple increment, then $FGT_{\alpha}(x; z) \leq FGT_{\alpha}(y;z)$.

	\vspace{2ex}
		\textsc{Monotonicity.} $\clx{M}$ satisfies monotonicity if, in addition to conforming to weak monotonicity, whenever $x$ is obtained from $y$ by a deprived increment among the poor, $FGT_{\alpha}(x;z) < FGT_{\alpha}(y;z)$.

	\vspace{2ex}
		\textsc{Dimensional monotonicity.} $\clx{M}$ satisfies dimensional monotonicity if whenever $x$ is obtained from $y$ by a dimensional increment among the poor, $FGT_{\alpha}(x; z) < FGT_{\alpha}(y;z)$.

	\vspace{2ex}
		\textsc{Nontriviality.} $\clx{M}$ is nontrivial if $FGT_{\alpha}$ achieves at least two different values.

	\vspace{2ex}
		\textsc{Normalization.} $\clx{M}$ is normalized if $0 \leq FGT_{\alpha}(y;z) \leq 1$ and these bounds are reachable.

	\vspace{2ex}
		\textsc{Weak transfer.} We say that $x$ is obtained from $y$ by \textit{averaging achievements among the poor} if $x = By$ for some bistochastic $N \times N$ matrix $B$ which satisfies $b_{ii} = 1$ if $i$ is non-poor. Thus $\clx{M}$ satisfies weak transfer if whenever $x$ is obtained from $y$ by averaging achievements among the poor, then $FGT_{\alpha}(x; z) \leq FGT_{\alpha}(y;z)$.

\begin{defn*}[Rearrangements] We say that $x$ is obtained from $y$ by a \textit{simple rearrangement among the poor} if there are two individuals $i, i'$ which are poor and such that for every $j$ it happens either $(x_{ij}, x_{i'j}) = (y_{i'j}, y_{ij})$ or $(x_{ij}, x_{i'j}) = (y_{ij}, y_{i'j})$, that is, the achievements of $i$ and $i'$ in dimension $j$ are either interchanged or not. Additionally, for every other person $i''$, $x_{i''j} = y_{i''j}$ for every $j$. If in addition the vectors $y_{i}$ and $y_{i'}$ were comparable by vector dominance but $x_{i}$ and $x_{i'}$ are not, then we say that $x$ is obtained from $y$ by an \textit{association decreasing rearrangement among the poor}. 
\end{defn*}
		
	\vspace{2ex}
		\textsc{Weak rearrangement.} $\clx{M}$ satisfies weak rearrangement if whenever $x$ is obtained from $y$ by an association decreasing rearrangement among the poor, then $FGT_{\alpha}(x; z) \leq FGT_{\alpha}(y ; z)$.

\newpage
%
%
	
\bibliographystyle{abbrvnat}
\bibliography{references}

\begin{thebibliography}{10}
\providecommand{\natexlab}[1]{#1}
\providecommand{\url}[1]{\texttt{#1}}
\expandafter\ifx\csname urlstyle\endcsname\relax
  \providecommand{\doi}[1]{doi: #1}\else
  \providecommand{\doi}{doi: \begingroup \urlstyle{rm}\Url}\fi

\bibitem[Alkire and Foster(2011)]{AlkireFoster11}
S.~Alkire and J.~Foster.
\newblock Counting and multidimensional poverty measurement.
\newblock \emph{Journal of Public Economics}, 95\penalty0 (7):\penalty0
  476--487, 2011.
\newblock \doi{10.1016/j.jpubeco.2010.11.006}.

\bibitem[Atkinson(1987)]{Atkinson87}
A.~B. Atkinson.
\newblock On the measurement of poverty.
\newblock \emph{Econometrica}, 55\penalty0 (4):\penalty0 749--764, 1987.
\newblock \doi{10.2307/1911028}.

\bibitem[Foster et~al.(1984)Foster, Greer, and Thorbecke]{FGTpaper}
J.~Foster, J.~Greer, and E.~Thorbecke.
\newblock A class of decomposable poverty measures.
\newblock \emph{Econometrica}, 52\penalty0 (3):\penalty0 761--766, 1984.
\newblock ISSN 00129682, 14680262.
\newblock URL \url{http://www.jstor.org/stable/1913475}.

\bibitem[Jehle(2011)]{JehleReny}
G.~A. Jehle.
\newblock \emph{Advanced microeconomic theory}.
\newblock Pearson, Harlow, 3. ed.. edition, 2011.
\newblock ISBN 9780273731917.

\bibitem[Lambert(2001)]{LambertPoverty}
P.~J. Lambert.
\newblock \emph{The distribution and redistribution of income : a mathematical
  analysis}.
\newblock Manchester University Press, 3rd ed.. edition, 2001.
\newblock ISBN 9780719057328.

\bibitem[Nussbaum and Sen(1993)]{SenFunctionings}
M.~Nussbaum and A.~Sen.
\newblock \emph{Capability and Well-Being}, chapter~2, pages 30--53.
\newblock Clarendon Press, Oxford, 1993.

\bibitem[P.~Chambers and D.~Miller(2007)]{BizarrePaper}
C.~P.~Chambers and A.~D.~Miller.
\newblock A measure of bizarreness.
\newblock \emph{Quarterly Journal of Political Science}, 5, 08 2007.
\newblock \doi{10.1561/100.00009022}.

\bibitem[Ravallion(2011)]{RavallionMPIBook}
M.~Ravallion.
\newblock \emph{On multidimensional indices of poverty}.
\newblock The World Bank, 2011.
\newblock \doi{10.1596/1813-9450-5580}.

\bibitem[Sen(1976)]{Sen76}
A.~Sen.
\newblock Poverty: An ordinal approach to measurement.
\newblock \emph{Econometrica}, 44\penalty0 (2):\penalty0 219--231, 1976.
\newblock \doi{10.2307/1912718}.

\bibitem[Sen(1979)]{Sen79}
A.~Sen.
\newblock Issues in the measurement of poverty.
\newblock \emph{The Scandinavian Journal of Economics}, 81\penalty0
  (2):\penalty0 285--307, 1979.

\end{thebibliography}

\end{document}